\documentclass[prl,twocolumn]{revtex4}

\usepackage{graphicx}

\begin{document}

\title{Optimized Non-Orthogonal Localized Orbitals for Electronic Structure
Calculations: Improved Linear Scaling Quantum Monte Carlo}
\author{Fernando A. Reboredo}
\email{reboredo1@llnl.gov}
\affiliation{Lawrence Livermore National Laboratory, Livermore, CA 94550}
\author{Andrew J. Williamson}
\email{williamson10@llnl.gov}
\affiliation{Lawrence Livermore National Laboratory, Livermore, CA 94550}

\begin{abstract}
  We derive an automatic procedure for generating a set of highly
  localized, non-orthogonal orbitals for linear scaling quantum Monte
  Carlo calculations.  We demonstrate the advantage of these orbitals
  in calculations of the total energy of both semiconducting and
  metallic systems by studying bulk silicon and the homogeneous
  electron gas.  For silicon, the improved localization of these
  orbitals reduces the computational time by a factor five and the
  memory by a factor of six compared to localized, orthogonal
  orbitals.  For jellium, we demonstrate that the total energy is
  converged for orbitals truncated within spheres with radii 7-8
  $r_s$, opening the possibility of linear scaling QMC calculations
  for realistic metallic systems.
\end{abstract}

\maketitle

In recent years, one of the most promising developments in the field
of electronic structure calculations has been the development of
algorithms whose cost grows as the first power of the system size.
Linear scaling variants of several electronic structure techniques
have been developed, including tight-binding\cite{galli2000}, density
functional theory
(DFT)\cite{li93,mauri94,kim95,hernandez95,galli2000,liu2000}, coupled
cluster\cite{schutz2001} and quantum Monte Carlo
(QMC)\cite{williamson2001,alfe2004}.  In all these approaches extended
Bloch orbitals, $\psi_{nk}$, are transformed into localized
Wannier-like orbitals. The speedup provided by the transformation to
localized orbitals depends on the extent to which the orbitals can be
localized and subsequently truncated.  Therefore, improved methods for
constructing localized orbitals have attracted intense attention in
recent years\cite{marzari1997,he2001,souza2001}.  Of particular
relevance to this paper is the recent demonstration that generalizing
from conventional orthonormal Wannier orbitals to non-orthogonal
orbitals\cite{galli2000,kim95,mauri94,liu2000,li93,hernandez95} can
provide increased localization and further accelerate linear scaling
DFT algorithms.

In QMC calculations, truncated, localized orbitals can be used to
introduce sparsity into the Slater determinant part of the trial wave
function.  As the calculation of the orbitals used to construct this
determinant is the dominant cost of QMC calculations, this
transformation yields a near linear scaling QMC algorithm.  In our
original approach to linear scaling QMC
calculations\cite{williamson2001}, the Slater determinant was
constructed from a set of orthonormal Wannier functions.  This choice
of orbitals produces a near linear scaling algorithm, which has
successfully been applied to calculations of the total energies and
optical gaps of a variety of semiconductor
systems\cite{williamson2002,puzder2003:2}.  However, this method
suffers from three main limitations: (i) It is only applicable to
systems where the Wannier functions decay rapidly (exponentially),
i.e., it works well for semiconductors and insulators, but it is not
applicable to metallic systems where orthonormal Wannier functions
decay polynomially\cite{he2001}.  (ii) The Wannier functions are
constructed via a unitary transformation of an input set of Bloch
functions.  This limits one to orthogonal functions, while a non-zero
Slater determinant requires only that the orbitals are linearly
independent.  (iii) Truncating a localized function reduces the volume
in which it needs to be evaluated, speeding up the calculation.
However, the Metropolis algorithm samples configurations of electron
coordinates from the many-body wavefunction, hence, some points are
sampled more frequently than others.  This knowledge of the
wavefunction is not included in the generation of the orthonormal
orbitals and hence the choice of orbitals is not optimal.

In this letter we derive and demonstrate the use of a {\em
  non-orthogonal} transformation of the Bloch orbitals that overcomes
the above limitations. This transformation is based on algorithms
developed for linear scaling DFT
calculations\cite{li93,mauri94,kim95,hernandez95,galli2000,liu2000}
and is designed to minimize a cost function associated with the total
number of orbital evaluations required in a linear scaling QMC
calculation.  For representative semiconductor systems, the orbitals
obtained from this non-orthogonal transformation are significantly
more localized and smoother than orthogonal Wannier functions, and can
typically be truncated in one sixth of the volume of the equivalent
orthogonal function without sacrificing accuracy.  This produces an
algorithm $\sim$5 times faster than previous linear scaling QMC
calculations\cite{williamson2001} and requiring one sixth of the
memory.  In addition, we demonstrate that while orthogonal Wannier
functions for metallic systems cannot be truncated within a practical
volume, non-orthogonal orbitals constructed via our procedure can be
truncated within a practical cutoff radius.

Our QMC calculations use a linear scaling\cite{williamson2001} version
of the CASINO\cite{casino} code with a standard Slater-Jastrow trial
wavefunction, $\Psi _{T}(\bf{R})$\cite{foulkes2001}.  The Slater
determinants are constructed from a set of truncated, localized linearly
independent orbitals $D_{ij}=\phi_i({\bf r}_j)\Theta_i({\bf r}_j)$,
where $\phi$ are are the non-orthogonal orbitals and $\Theta$ are the
truncation functions.  In principle, one can optimize the shape of the
truncation functions, however, for the systems studied here, we find
that spherical step functions are a simple and stable solution where,
\begin{eqnarray}
\Theta ^{i}(\mathbf{r}) &=&1\;\;,\;\;|\mathbf{r-R_{i}}|<R_{i}^{cut} 
\nonumber  \label{theta} \\
&=&0\;\;,\;\;|\mathbf{r-R_{i}}|>R_{i}^{cut}\;\;\;.
\end{eqnarray}
The truncation functions $\Theta ^{i}$ are defined by two parameters,
the cutoff radii $R_{i}^{cut}$ and the centers $\mathbf{R_{i}}$.
These parameters are optimized iteratively using a procedure designed
to minimize the computational cost of the QMC calculation. The
non-orthogonal orbitals, $\phi_i$, associated with each $\Theta^{i}$
are obtained during the iterative process.

The computational cost of a typical QMC calculation is proportional to
the number of orbital evaluations required to construct the Slater
determinant for each configuration of electron coordinates, ${\bf
  R}=({\bf r}_1,{\bf r}_2...  {\bf r}_N)$.  The cost is therefore the
product of the probability, $|\Psi _{T}({\bf R})|^{2}$, of sampling a
given configuration, ${\bf R}$ and the cost of evaluating each of the
non-zero elements in the Slater determinant produced by that
configuration.  For each element, $\phi_i({\bf r}_j)$, if ${\bf r}_j$
falls within the truncation function, $\Theta_i$, this adds 1 to the
cost, i.e.
\begin{equation}\label{cost}
\mbox{Cost}  =  \int d\mathbf{R}\;|\Psi _{T}(\mathbf{R})|^{2}\sum_{ij}^{N}\Theta_i({\bf r}_j) \;\;\;.
\end{equation}
By integrating out all but one electron coordinates, Eq.(\ref{cost})
can be expressed in terms of the density $\rho({\bf r})$ as
\begin{equation}\label{cost2}
\mbox{Cost}  =  \sum_i \int d{\bf r} \; \rho({\bf r}) \Theta_i({\bf r}) \; . 
\end{equation}
We find a satisfactory minimum of Eq. (\ref{cost2}) by starting from
an initial choice of $\Theta_i$ and iteratively updating first the
cutoff radii $R_{i}^{cut}$ and then the centers ${\bf R}_i$.

\begin{figure}[tbp]
  \includegraphics[width=0.95\linewidth,clip=true]{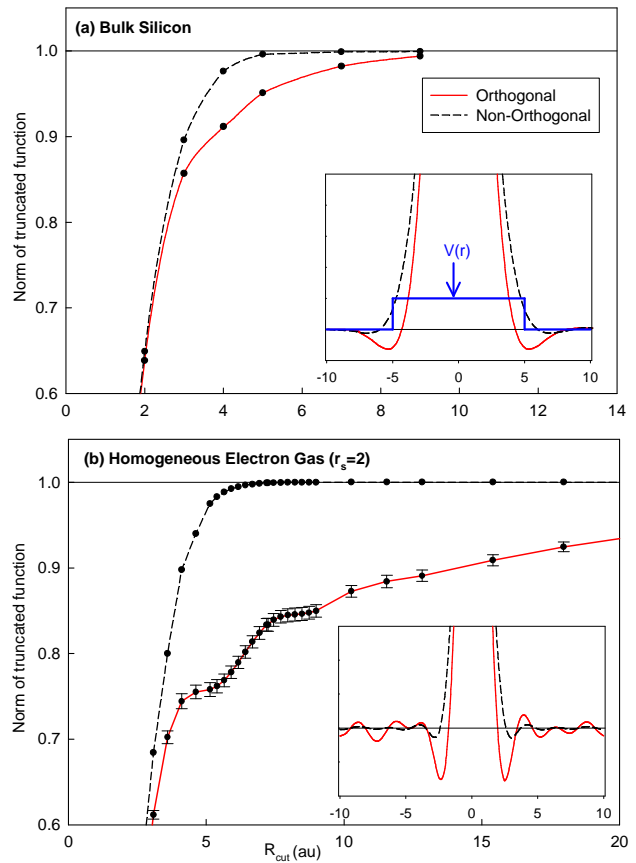}
\caption{(Color) Comparison of the norm of orthogonal and
  non-orthogonal localized orbitals in (a) bulk silicon and (b) a HEG
  at the same $r_s$. The error bars show the spread in norm between
  the states. The insets compare the shape of the orthogonal and
  non-orthogonal localized orbitals.}
\label{norm_fig}
\end{figure}

\noindent {\em (i) Generating Optimal Non-Orthogonal Orbitals and Cutoff 
  Radii}: Each truncation function, $\Theta_i$, can be considered as a
potential, $\hat{V}$, acting on the Hilbert space of Bloch orbitals.
In the inset to Fig~\ref{norm_fig}a this potential, $\hat{V}$, is
shown with a blue line.  If one constructs the matrix elements of the
Bloch orbitals with this potential,
$V_{jk}^{i}=<\phi_j^{Bloch}|\Theta^i({\bf r})|\phi_k^{Bloch}>$, then
the eigenstate $\phi_i$ of $V^{i}$ with the largest eigenvalue is the
most localized state within the truncation region.  This is the
orbital with the maximum truncated norm, $X$, defined as
\begin{equation}\label{xdef}
X = \int d{\bf r} |\phi_i({\bf r})|^2 \Theta_i({\bf r}) \;\;\;.
\end{equation}
Increasing $R_{i}^{ cut}$ increases the above value of $X$, reducing
the resulting truncation error in the QMC calculation, but also
increases the computational cost in Eq.(\ref{cost2}).  Therefore, we
adjust the cutoff radius $R_{i}^{cut}$ to achieve a target norm, e.g.
$X=0.999$.  Repeating this diagonalization procedure for each
truncation function $\Theta^i$ generates an associated set of
non-orthogonal orbitals $\{\phi\}$.  This procedure for generating a
set of non-orthogonal orbitals associated with a set of truncation
regions is similar to those adopted in linear scaling density
functional calculations\cite{mauri94,kim95,galli2000,liu2000} and
recently in a QMC calculation of MgO\cite{alfe2004}.  Next, we extend
this procedure to automatically optimize the centers of the truncation
functions for systems where they cannot be guessed {\em a priori}.

\noindent {\em (ii) Updating the Truncation Centers:}
The cost function in Eq. (\ref{cost2}) can be rewritten as 
\begin{equation}\label{cost3}
\mbox{Cost}  =  N X + \sum_i \int d{\bf r} \left[ \rho({\bf r}) - |\phi_i({\bf r})|^2 \right] \Theta_i({\bf r})\;\;\;,
\end{equation}
where $N$ is the number of orbitals and $X$ is defined in
Eq.(\ref{xdef}).  The first term in Eq.(\ref{cost3}, $NX$, cannot be
reduced without losing accuracy.  Therefore the only way to reduce the
computational cost is to minimize the second term in Eq.(\ref{cost3}
by placing the truncation centers where $ \rho({\bf r}) - |\phi_i({\bf
  r})|^2$ is minimum.  Since $\rho({\bf r}) \geq |\phi_i({\bf r})|^2$,
this is minimized in regions where $\phi_i$ is most localized and
therefore closest to $\rho$.  Therefore for the next iteration, we
move the truncation centers towards the center of mass of the
$|\phi_i({\bf r})|^2$ for the current iteration.  To ensure linear
independence, we orthogonalize the set $\{ \phi\}$ with a polar
decomposition before calculating this center of mass.

This updated set of truncation functions, $\Theta^i$, with new
centers, ${\bf R}_i$, are then used to generate a new set of
non-orthogonal orbitals using the procedure in (i) above and the
process is repeated.  Starting from a random choice of centers 10-15
iterations are typically required to find a minimum of Eq.
(\ref{cost2}) and to converge the centers.  If one uses a good
starting set of centers, such as the centers of maximally localized
Wannier functions\cite{marzari1997}, the $\Theta_i$ converge in 1 or 2
iterations.

To analyze the properties of these new non-orthogonal orbitals we
first compare their localization and decay properties with an
equivalent set of orthogonal orbitals.  We then examine the
convergence of the total energy in quantum Monte Carlo calculations
using these orbitals.  Comparisons are made for the prototypical
semiconductor and metal systems, silicon and the homogeneous electron
gas.

\begin{figure}[tbp]
\includegraphics[width=0.95\linewidth,clip=true]{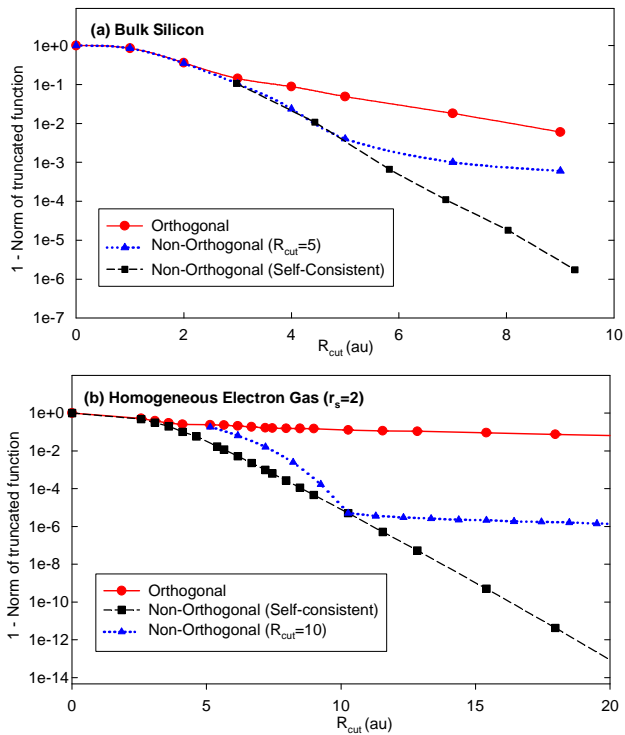}
\caption{(Color) Comparison of the decay of orthogonal , self
  consistent, non-orthogonal and fixed radius, non-orthogonal
  localized orbitals in (a) bulk silicon and (b) HEG ($r_{s}=2$). }
\label{decay_fig}
\end{figure}

In Fig.~\ref{norm_fig} the norm within a spherical truncation region
of the orthogonal and non-orthogonal localized orbitals are compared
as a function of $R_{cut}$. Fig.~\ref{norm_fig}(a) compares orbitals
constructed for bulk silicon. The input states were obtained from a 64
atom LDA calculation\cite{GP}, using a norm-conserving Hamann
pseudopotential and a 35 Ry cutoff. The non-orthogonal orbitals were
constructed using the iterative procedure described above, where the
desired norm, $X$, was varied from 0.5 to 0.99999 to construct the
plot. The orthogonal states were obtained by performing a final polar
decomposition orthogonalization step.  Due to symmetry all the
non-orthogonal orbitals are equivalent.  Within a given radius, the
non-orthogonal orbitals contain significantly more charge, e.g.
99.9\% of the norm is contained within a sphere of radius 5.5~au
compared to orthogonal orbitals which require an 11 au sphere to
capture the same charge.  The origin of this dramatically improved
localization is shown in the inset to Fig.~\ref{norm_fig}(a) which
shows a line plot through the center of the $R_{cut}=2$ orbitals.
While the non-orthogonal states decay smoothly to zero with minimal
oscillations, the orthogonal orbitals oscillate around zero for $>5$
au after initially crossing zero to maintain orthogonality between
states.  While the amplitude of these oscillations is small compared
to the central peak, the $r^{2}$ prefactor leaves a significant amount
of charge in these oscillations.

Comparing the orthogonal orbitals shown in Fig.~\ref{norm_fig}(a) with
maximally localized (MLW) orbitals constructed according to
Ref.\cite{marzari1997}, which essentially finds the localized
eigenstates of the $e^{i2\pi \mathbf{r}/L}$ operator, we find the
centers of our non-orthogonal and orthogonal states are identical to
the MLW function centers due to symmetry.  Additionally, the shape and
norm convergence of our orthogonal states is almost identical to the
MLW functions. It therefore appears that the shape of
\emph{orthogonal} localized orbitals is relatively insensitive to the
choice of operator used to localize the states.

Figure~\ref{norm_fig}(b) shows orthogonal and non-orthogonal orbitals
constructed for the HEG with $r_{s}=2$ (same as silicon). The input
states were the lowest 1935 plane waves in a 50 au cubic box.  The
norm of the orthogonal orbitals slowly approaches 1.0 as the radius is
increased, as would be expected given the slow polynomial decay of
orthogonal orbitals in metallic systems\cite{he2001}.  In contrast,
the non-orthogonal orbitals rapidly approach 1.  For example 99.9\% of
the norm of the non-orthogonal orbitals is contained within a sphere
of radius 7~au, while even the largest sphere inscribed within the
supercell (25 au radius) contains only 94\% of the norm of the
orthogonal orbitals.  Note, the non-orthogonal orbitals are still less
localized than those in silicon, where 99.9\% of the norm is contained
within a sphere of radius 5.5~au compared to the 7~au required for
jellium.  As in silicon, the inset plot shows pronounced, long range
oscillations in the orthogonal orbitals and a much smoother decay of
the non-orthogonal orbitals with minimal oscillation.

Figure~\ref{decay_fig} compares the truncated decay of orthogonal and
non-orthogonal localized orbitals for bulk silicon and the HEG.  This
is equivalent to the decay of the trace of the non-orthogonal density
matrix\cite{li93,hernandez95,galli2000}.  Here we define the decay as
1 minus the norm contained within a sphere of radius $R$.  As expected
from Fig.~\ref{norm_fig}, Fig.~\ref{decay_fig} shows that the
non-orthogonal orbitals (black dashed lines) decay more rapidly than
the equivalent orthogonal orbitals (red solid lines).
Figure~\ref{decay_fig} also illustrates that to obtain maximum
localization within a given volume, the non-orthogonal orbitals must
be adjusted consistently with $R_{cut}$\cite{note1}.  The blue dotted
line in Fig.\ref{decay_fig}b shows the decay of a non-orthogonal
orbital optimized to be maximally localized in a $\Theta$ function
with $R_{cut}=10$.  The truncated norm was then evaluated for a range
of $R$ values while keeping the orbital fixed.  For $R_{cut}=10$ the
``self-consistent'' and fixed non-orthogonal orbitals are identical.
For all other values of $R_{cut}$ , the orbital optimized with
$R_{cut}=10$ (blue,dotted) is no longer the optimal orbital for that
choice of $\Theta$ function and it therefore has a lower truncated
norm.
\begin{figure}[tbp]
\includegraphics[width=0.95\linewidth,clip=true]{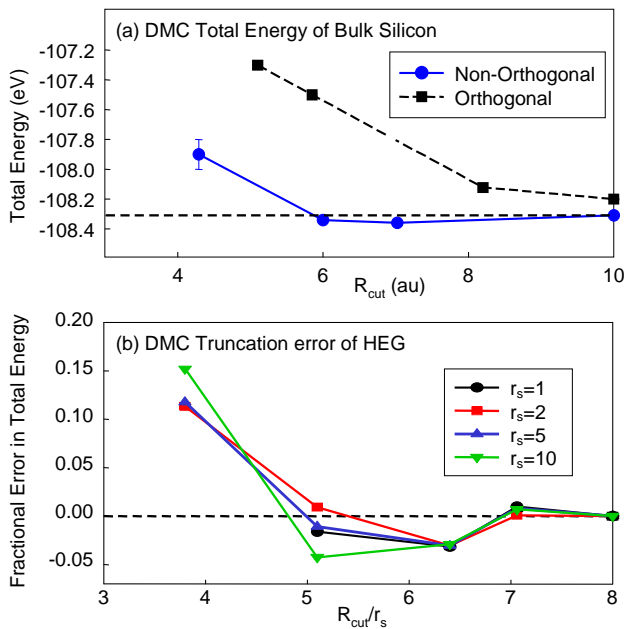}
\caption{(Color) (a) Convergence of DMC total energy of bulk silicon with
  truncation radius for orthogonal and non-orthogonal orbitals, (b)
  Convergence of DMC total energy of the HEG as a function of
  $R_{cut}/r_s$}
\label{dmc_fig}
\end{figure}

Our previous work with orthogonal, truncated, localized
orbitals\cite{williamson2001} indicated that the norm of these
orbitals was a good predictor of the truncation error in a QMC
calculation the total energy.  For silicon we found that a truncation
region large enough to capture 99.9\% of the norm was sufficient to
produce a converged total energy.  On this basis, the improved decay
properties of the non-orthogonal orbitals shown in
Figs.~\ref{norm_fig} and \ref{decay_fig} suggest that these orbitals
can be used to perform QMC calculations with smaller truncation radii
than those used for previous orthogonal orbitals, without sacrificing
accuracy.  However, the decay of the localized orbitals in a given
representation does not predict the truncation error in the density
matrix and how electron-electron correlation will be affected.
Therefore, to fully evaluate the properties of these non-orthogonal
orbitals we have performed VMC and DMC total energy calculations of
bulk silicon and the HEG to compare the convergence of the total
energy with the truncation radii of the localized orbitals.

Figure~\ref{dmc_fig}(a) compares the convergence of the DMC total
energy of the same 64 atom, bulk silicon system shown in
Figs.~\ref{norm_fig}(a) and Fig.~\ref{decay_fig} (a), using orthogonal
and non-orthogonal input orbitals. It shows that the DMC energy
converges more rapidly using non-orthogonal orbitals.  To converge the
total energy to with 0.01 eV per atom using orthogonal orbitals
required $R_{cut}=11$ au\cite{williamson2001} while using
non-orthogonal orbitals, equivalent accuracy can be obtained with
$R_{cut}=6$ au. This results in a factor of 5 increase in speed and a
factor of 6 reduction in memory.

Figure~\ref{dmc_fig}(b) compares the convergence with $R_{cut}$ of the
DMC total energy of a homogeneous electron gas with $r_{s}=1,2,5$ and
10.  In the HEG, the non-orthogonal orbitals for all $r_{s}$ values
can be obtained by scaling the $r_{s}=1$ orbital.  The kinetic energy
scales as $r_s^{-2}$. To enable us to plot all values of $r_s$ on the
same plot, we rescale both axes and plot the fractional DMC truncation
error, defined as
$\mbox{Error}(R_{cut})=[E(R_{cut})-E_{\infty}]/E_{\infty}$ as a
function of $R_{cut}/r_{s}$. After this rescaling the convergence
plots for each value of $r_{s}$ fall on a similar curve.  Note, the
negative truncation error around $R_{cut}/r_{s}=6$ resulting from a
loss of kinetic energy, due to abrupt truncation of the orbitals.
This curve shows that the total DMC energy is approximately converged
for truncation radii of $7-8 r_{s}$.  These converged values are in
excellent agreement with the original values from Ceperley and
Alder\cite{ceperley1980}.  Therefore, while the slower polynomial
decay of the density matrix of metallic systems requires a larger
truncation radius to converge the total energy than for semiconductors
with equivalent density, the above procedure for generating
non-orthogonal orbitals does allow the localized orbitals for metallic
systems to be truncated in a practical volume for linear scaling
calculations.  In addition, the above procedure for generating these
non-orthogonal orbitals does not require the high symmetry of the HEG
and therefore this approach could be equally applied to linear scaling
DMC calculations of realistic metallic systems.

In conclusion, we derive a simple, automatic pre-processing procedure
for generating non-orthogonal localized orbitals which minimize the
total computational cost of linear scaling QMC calculations. We
demonstrate the application of these orbitals to DMC calculations of
the prototypical semiconductor and metallic systems, silicon and the
HEG.  We anticipate that these orbitals may also have applications in
alternative electronic structure techniques such as DFT which also
utilize localized orbitals to generate linear scaling.

The authors would like acknowledge G. Galli, R. Needs, R. Hood and D.
Prendergast for helpful discussions and comments.  This work was
performed under the auspices of the U.S. Department of Energy by the
University of California, Lawrence Livermore National Laboratory under
contract No. W-7405-Eng-48.


\begin{thebibliography}{18}
\bibitem{note1}Adjusting $R_{cut}$ introduces minimal cost for QMC
  calculations as the radii are adjusted once before the
  calculation.
\end{thebibliography}

\end{document}